\documentstyle[12pt,fleqn]{article}
\renewcommand{\theequation}{\arabic{section}.\arabic{equation}}
\renewcommand{\thesection}{\arabic{section}.}

\catcode`@=11 \@addtoreset{equation}{section} \catcode`@=12
\mathchardef\SGamma="7100

\begin{document}

\title{{\bf Nonperturbative late time asymptotics for
heat kernel in gravity theory}}
\date{}
\author{A.O.Barvinsky$^{1}$, Yu.V.Gusev$^{2}$, V.F.Mukhanov$^{3}$
and D.V.Nesterov$^{1}$\\
$^{1}${\em Theory Department, Lebedev Physics Institute,}\\
{\em Leninsky prospect 53, Moscow 117924, Russia}\\
$^{2}${\em  Pacific Institute for Mathematical Sciences and} \\
{\em Department of Physics and Astronomy,}\\
{\em University of British Columbia,}\\
{\em 6224 Agricultural Rd., B.C. Canada V6T 1Z1}\\
$^{3}${\em Sektion Physik, LMU, Theresienstr.37, Munich, Germany}}
\maketitle

\begin{abstract}
Recently proposed nonlocal and nonperturbative late time behavior
of the heat kernel is generalized to curved spacetimes. Heat
kernel trace asymptotics is dominated by two terms one of which
represents a trivial covariantization of the flat-space result and
another one is given by the Gibbons-Hawking integral over
asymptotically-flat infinity. Nonlocal terms of the effective
action generated by this asymptotics might underly long-distance
modifications of the Einstein theory motivated by the cosmological
constant problem. New mechanisms of the cosmological constant
induced by infrared effects of matter and graviton loops are
briefly discussed.
\end{abstract}

\section{Introduction}
\hspace{\parindent} In this paper we continue the studies of
nonperturbative infrared behavior in field-theoretical models
initiated in \cite{nnea}. In contrast to ultraviolet properties
incorporating renormalized coupling constants -- coefficients of
local invariants in the action, infrared behavior manifests itself
in nonlocal structures responsible for long-distance phenomena.
The growing interest in such phenomena, especially in context of
the gravitational theory, arises due to recent attempts of
resolving the cosmological constant problem by means of
long-distance modifications of Einstein theory. Moreover, these
modifications often call for nonperturbative treatment in view of
the nonlinear aspects of van Damm-Veltman-Zakharov discontinuity
problem \cite{VVDZ} and the presence of a hidden nonperturbative
scale in gravitational models with extra dimensions \cite{scale}.
On the other hand, nonlocalities also (and, moreover, primarily)
arise in virtue of fundamental quantum  effects of matter and
graviton loops which, for instance, can play important role in
gravitational radiation theory \cite{MWZ,vilkov} and cosmology
\cite{Woodard}. Therefore, they can successfully compete with
popular phenomenological mechanisms of infrared modifications,
induced, say, by braneworld scenarios with extra dimensions
\cite{GRS,DGP} or other models \cite{WoodMOND}. This makes
nonperturbative analysis of nonlocal quantum effects very
interesting and promising.

Nonlocal quantum effects can be described by the Schwinger-DeWitt
proper time method based on the heat kernel
    \begin{eqnarray}
    K(s|\,x,y)=\exp\Big[s\,F(\nabla)\Big]\,
    \delta(x,y),                 \label{1.0}
    \end{eqnarray}
which is a main building block of Feynman diagrams with the
inverse propagator $F(\nabla)$ -- the differential operator of
field disturbances on some matter and gravitational field
background \cite{DeWitt,PhysRep}. The infrared physics is then
determined by the late time behavior of
$K(s)=\exp\big[s\,F(\nabla)\big]$ and by its functional trace that
generates the one-loop effective action \cite{DeWitt,PhysRep,nnea}
    \begin{eqnarray}
    &&\SGamma_{\rm one-loop}\equiv
    \frac12\,{\rm Tr}\,\ln\,F(\nabla)=
    -\frac12\,\int_0^\infty\frac{ds}s\,
    {\rm Tr}\,K(s),                \label{1.000}\\
    &&{\rm Tr}\,K(s)
    \equiv\int dx\,K(s\,|\,x,x). \label{1.00}
    \end{eqnarray}
In particular, nonlocal terms of $\SGamma_{\rm one-loop}$ arise as
a contribution of the upper limit in the proper-time integral
(\ref{1.000}), which makes the late time asymptotics of ${\rm
Tr}\,K(s)$ most important\footnote{Here the effective action is
defined in Euclidean space with positive-signature metric. Its
application in physical spacetime with Lorentzian signature is
based on analytic continuation methods which range from a conventional
Wick rotation in scattering theory (for in-out matrix elements) to
a special retardation prescription in a wide class of problems for
a mean field (in-in expectation value) \cite{CPTI,nonlocal}.
These methods nontrivially apply to nonlocal terms and extend
from the usual perturbation theory to its partial resummation
corresponding to the nonperturbative technique of the present
work.}.

The heat kernel trace, including its late time asymptotics, was
first studied within the covariant nonlocal curvature expansion in
\cite{CPTI,CPTII,CPTIII,asymp}. Then its nonperturbative
asymptotics was obtained in \cite{nnea} for a particular case of
flat spacetime and used to derive new nonlocal and essentially
nonlinear terms in the effective action (\ref{1.000}). These terms
represent a generalization of the logarithmic Coleman-Weinberg
potential to the case of nonconstant fields vanishing at infinity
and, thus, generating special type of nonlocal behavior. The goal
of this paper is to generalize it to curved spacetime with generic
asymptotically-flat geometry. In the next section we begin by
formulating our main results after briefly recapitulating the
setting of the problem and conclusions of \cite{nnea}.

\section{The setting of the problem and main results}
\hspace{\parindent} In \cite{nnea} it was shown that the heat
kernel of the differential operator $F(\nabla)$ with generic
potential $V(x)$ in flat Euclidean (positive signature) spacetime
with $d$ dimensions,
    \begin{equation}
    F(\nabla)=\Box-V(x),\,\,\,\Box=\nabla^\mu\nabla_\mu,  \label{1.1}
    \end{equation}
has a nonperturbative in potential late time expansion
    \begin{equation}
    K(s|\,x,y)=\frac1{(4\pi s)^{d/2}}\,
    \exp\left(-\frac{|x-y|^2}{4s}\right)\,
    \left[\,\Phi (x)\,\Phi (y)
    +O\left(\,\frac1s\,\right)\,\right],\,\,\,
    s\rightarrow\infty .                  \label{1.1a}
    \end{equation}
Its leading order behaviour is determined in terms of a special function
$\Phi(x)$ -- the solution of the homogeneous equation with unit boundary
condition at infinity
    \begin{eqnarray}
    &&F(\nabla)\,\Phi(x)=0, \nonumber\\
    &&\Phi(x)\to 1, \,\,\,\,|x|\to\infty.  \label{1.5a}
    \end{eqnarray}
This solution can be represented in a closed form in terms of the Green's
function of (\ref{1.1}) with zero boundary conditions at $|x|\to\infty$,
$G(x,y)$,
    \begin{equation}
    \Phi (x)=1+\frac{1}{\Box-V}\,V(x)\equiv
    1+\int dy\,G(x,y)\,V(y).  \label{1.5}
    \end{equation}

As a byproduct of this result it was also shown that the
functional trace of this heat kernel has an asymptotic
$1/s$-expansion beginning with
    \begin{eqnarray}
    {\rm Tr}\,K(s)=\frac1{(4\pi s)^{d/2}}
    \int dx\,\left[\,-s\,V\,\Phi
    -2\,\nabla_\mu\Phi\,\frac1{\Box-V}\,
    \nabla^\mu\Phi+1
    +O\left(\frac1s\right)\,\right].        \label{1.2}
    \end{eqnarray}
Important property of this functional trace that was understood in
\cite{nnea} is that this expansion cannot be obtained directly by
integrating the coincidence limit of the expansion (\ref{1.1a}).
This happens because the latter is not uniform in $|x|\to\infty$
and, therefore, yields erroneous contribution when integrating
over infinite spacetime. This explains, in particular, why the
leading behaviour of (\ref{1.2}) is $O(s/s^{d/2})$ in contrast to
that of (\ref{1.1a}), $O(1/s^{d/2})$. Nevertheless, (\ref{1.2})
can be recovered from (\ref{1.1a}) by functionally integrating the
variational equation
    \begin{equation}
    \frac{\delta \,{\rm Tr}K(s)}{\delta V(x)}
    =-sK(s|x,x).                               \label{1.2a}
    \end{equation}
The leading term of the asymptotic expansion (\ref{1.2}) was
obtained in \cite{nnea} exactly by this procedure. This result was
also verified in \cite{nnea} by a direct summation of the
covariant perturbation theory developed for the heat kernel trace in
\cite{CPTII}. The subleading term of (\ref{1.2}) was derived
entirely by the second method, because the corresponding
term of the heat kernel (\ref{1.1a}) was not yet known.

In this paper we generalize the results of \cite{nnea} to the case
of the scalar operator (\ref{1.1}) in curved spacetime with the
covariant d'Alembertian (Laplacian in Euclidean space)
defined with respect to generic asymptotically
flat metric $g_{\mu\nu}(x)$
    \begin{equation}
    \Box=g^{\mu\nu}(x)\nabla_\mu\nabla_\nu=
    \frac1{g^{1/2}(x)}
    \frac\partial{\partial x^\mu}\,g^{1/2}(x)
    g^{\mu\nu}(x)\frac\partial{\partial x^\nu}.   \label{1.3}
    \end{equation}
We obtain the heat kernel in the first two orders of
the $1/s$-expansion. It has the form
    \begin{eqnarray}
    &&K(s|\,x,y)=\frac1{(4\pi s)^{d/2}}\,
    \exp\left[-\frac{\sigma(x,y)}{2s}\right]\nonumber\\
    &&\qquad\qquad\qquad\qquad\qquad
    \times\left\{\,\Phi (x)\,\Phi (y)
    +\frac1s\,\Omega_1(x,y)+
    O\left(\,\frac1{s^2}\right)\,
    \right\}\,g^{1/2}(y),                  \label{1.4}
    \end{eqnarray}
where $\sigma(x,y)$ is the the world function -- one half of the
geodesic distance between the points $x$ and $y$. The leading order is
again defined by the function (\ref{1.5}) which is
determined in terms of the Green's function,
    \begin{eqnarray}
    G(x,y)=\frac1{F(\nabla)}\delta(x,y)    \label{1.4a}
    \end{eqnarray}
of the curved space operator $F(\nabla)$ with the covariant
d'Alembertian (\ref{1.3}) and with Dirichlet boundary conditions
at infinity\footnote{We define the $\delta(x,y)$-function as a
scalar with respect to the first argument $x$ and as a density of
unit weight with respect to the second one -- $y$. Correspondingly
the heat kernel defined by Eq.(\ref{1.0}) and the kernel of the
Green's function $G(x,y)$ have the same weights of their
arguments. This asymmetry in $x$ and $y$ explains the presence of
the factor $g^{1/2}(y)$ in (\ref{1.4}) and a biscalar nature of
$\Omega_1(x,y)$.}. The subleading term is more complicated, and
the expression for $\Omega_1(x,y)$ is presented in Sect.4 below.

We also derive the asymptotics of the functional trace of the
heat kernel corresponding to (\ref{1.4})
    \begin{eqnarray}
    {\rm Tr}K(s)=\frac1{(4\pi s)^{d/2}}\,
    \left\{\,s\,W_0+W_1
    +O\left(\,\frac1s\,\right)\,\right\}.   \label{1.6}
    \end{eqnarray}
The leading term here turns out to be a covariantized (curved
space) version of the same term in the flat-space trace
(\ref{1.2}) plus the surface integral of the local function of
metric and its first-order derivatives at spacetime infinity,
    \begin{eqnarray}
    &&W_0=-\int dx\,g^{1/2}\,V\,\Phi(x)
    +\frac16\,\Sigma[\,g_\infty\,],    \label{1.7}\\
    &&\Sigma[\,g_\infty\,]
    =\int\limits_{|x|\to\infty}\!
    d\sigma^\mu\;\delta^{\alpha\beta}
    \Big(\partial_\alpha
    g_{\beta\mu}-\partial_\mu g_{\alpha\beta}\Big). \label{1.7a}
    \end{eqnarray}
This surface integral over the sphere of radius $|x|\to\infty$ is
written here in cartesian coordinates and involves only the
flat-space asymptotics of the metric
    \begin{eqnarray}
    g^\infty_{\mu\nu}(x)\equiv
    g_{\mu\nu}(x)\,\Big|_{\,|x|\to\infty}
    =\delta_{\mu\nu}
    +O\left(\frac1{|x|^{d-2}}\right).  \label{1.7b}
    \end{eqnarray}
Its covariant version in the form of the Gibbons-Hawking surface
integral of the extrinsic curvature of the boundary is discussed
in Sect.6. We also demonstrate that the subleading term $W_1$
confirms the result of the direct summation of perturbation series
in potential \cite{nnea} in the flat-space case (\ref{1.2}).

The organization of the paper is as follows. In Sect.3 we derive
the technique of recurrent equations for the coefficients of the
$1/s$-expansion of the heat kernel and discuss the peculiarities
of setting their boundary value problem. We apply this technique
in the leading order of the asymptotic expansion and derive the
$V$-dependent part of the algorithm (\ref{1.7}). This is achieved
by functionally integrating the variational equation (\ref{1.2a})
in the leading order of $1/s$-expansion. In Sect. 4 this technique
is extended to the subleading order, and it is shown how it
reproduces in flat spacetime the third term of (\ref{1.2}) -- the
result obtained in \cite{nnea} by tedious summation of nonlocal
perturbation series. In Sect. 5 we perform a major check on the
correctness of the metric dependence of the late-time asymptotics
of ${\rm Tr}\,K(s)$. The functional integration of Eq.(\ref{1.2a})
with respect to a potential determines ${\rm Tr}\,K(s)$ only up to
an arbitrary functional of the metric independent of $V$. So we
derive the metric variational equation analogous to (\ref{1.2a})
and show that the bulk part of $W_0$ in (\ref{1.7}) exactly
satisfies this equation. However, local functional derivative with
respect to $g_{\mu\nu}(x)$ in the bulk (that is for finite $|x|$)
does not feel the asymptotic surface term of the form
(\ref{1.7a}), so in order to establish the latter we compare in
Sect. 6 the nonperturbative asymptotics of ${\rm Tr}\,K(s)$ with
its covariant curvature expansion of \cite{CPTIII,asymp}. This
comparison confirms the bulk structure of the algorithms
(\ref{1.7}) and also fixes the additional surface integral -- the
Gibbons-Hawking term (\ref{1.7a}). As a byproduct of this
procedure we establish a new representation for this surface term
in the form of the bulk integral of the {\em nonlocal} Lagrangian
which is expanded in covariant curvature series and explicitly
independent of such auxiliary quantities as extrinsic curvature of
the boundary. In the concluding section we list the omissions of
the proposed formalism, the prospects of its extension beyond the
leading order and its generalizations to spacetimes with other
than asymptotically-flat boundary conditions. We also briefly
discuss the status of the cosmological constant induced by
nonperturbative effective action which originates from the late
time asymptotics of the above type. In the appendix we present the
variational formalism used in the subleading order of the late
time expansion.

\section{Heat kernel and heat kernel trace at late times}
\hspace{\parindent}
To find late time asymptotics of the heat kernel in curved space
we use the ansatz
    \begin{eqnarray}
    &&K(s|\,x,y)=\frac1{(4\pi s)^{d/2}}\,
    \exp\left[-\frac{\sigma(x,y)}{2s}\right]\,
    \Omega(s|\,x,y)\,g^{1/2}(y),               \label{2.1}\\
    &&\Omega(s|\,x,y)=
    \Omega_0(x,y)+\frac1s\,\Omega_1(x,y)
    +O\left(\,\frac1{s^2}\,\right).              \label{2.3}
    \end{eqnarray}
Here $\sigma(x,y)$ is a world function -- one half of the geodesic
distance between the points $x$ and $y$ -- satisfying the equation
    \begin{eqnarray}
    \frac12\,g^{\mu\nu}(x)\,\nabla_\mu\sigma(x,y)\,
    \nabla_\nu\sigma(x,y)=\sigma(x,y).
    \end{eqnarray}
This ansatz is motivated by the small time limit of the heat
kernel in which $\Omega(s|\,x,y)$ has a regular Schwinger-DeWitt
expansion in powers of $s$, $\Omega(s|\,x,y)=
\Delta^{1/2}(x,y)\,[\,1+O(s)\,]$, where the overall factor
$\Delta^{1/2}(x,y)=g^{-1/4}(x)\,[\det
\partial^x_\mu\partial^y_\nu\sigma(x,y)]\,g^{-1/4}(y)$ is
the (dedensitized) Pauli-Van Vleck-Morette determinant.

As we will see in what follows, disentangling of $\Delta^{1/2}(x,y)$
as a separate factor in (\ref{2.1}) is not useful for the purposes
of late time expansion. However, the quantity is rather
important and related to a serious simplifying assumption which
underlies our results. The assumption we make is the absence of
focal points in the congruence of geodesics determining the
world function $\sigma(x,y)$. We assume that
for all pairs of points $x$ and $y$, $\Delta(x,y)\neq 0$,
which guarantees that $\sigma(x,y)$ is globally and uniquely
defined on the asymptotically-flat spacetime in question. This
assumption justifies the ansatz (\ref{2.1})-(\ref{2.3}) which
should be globally valid because the coefficients of the
expansion (\ref{2.3}) will satisfy elliptic boundary-value
problems with boundary conditions at infinity. This
requirement is independent of the asymptotic flatness because
the presence of caustics in the geodesic flow, $\Delta(x,y)=0$,
might depend on local properties of the gravitational field,
unrelated to its long-distance behavior. Roughly, the
gravitational field should not be too strong to
guarantee the geodesic convexity of the whole spacetime.
This assumption might be too strong to incorporate physically
interesting situations, but we believe that the main result
will survive the presence of caustics (though, maybe by the
price of additional contributions which are essentially
nonperturbative and go beyond the scope of this paper)\footnote{
This hope is based on a simple fact that the leading order
of the $1/s$-expansion -- the primary object of this paper --
is not sensitive to the properties
of the world function at all (see Eq. (\ref{2.4}) below, which
does not involve $\sigma(x,y)$).
Beyond this order the main object of interest, ${\rm Tr}\,K(s)$,
involves the coincidence limit of the world function
$\sigma(x,x)=0$, while its asymptotic coefficients in
(\ref{2.3}) nonlocally depend on global geometry and
can acquire from caustics additional contributions analogous
to those of multiple geodesics connecting the points $x$ and $y$
beyond the geodesically convex neighborhood \cite{camporesi}.}.

Substituting the ansatz (\ref{2.1}) in the heat equation
    \begin{eqnarray}
    \frac\partial{\partial s}K(s|\,x,y)=
    F(\nabla_x)\,K(s|\,x,y)
    \end{eqnarray}
one obtains the equation for the unknown function $\Omega(s|\,x,y)$
    \begin{eqnarray}
    \frac{\partial\Omega}{\partial s}
    +\frac1s\left(\sigma^\mu\nabla_\mu
    +\frac12\,\Box\,\sigma
    -\frac{d}2\,\right)\Omega
    =F(\nabla)\,\Omega,                        \label{2.2}
    \end{eqnarray}
where $\sigma^\mu\equiv\nabla^\mu_x\sigma(x,y)$, and
$\Box\,\sigma\equiv\Box_x\sigma(x,y)$.

Assuming the validity of the $1/s$-expansion (\ref{2.3}) for
$\Omega(s|\,x,y)$ at $s\to\infty$ (which follows, in particular,
from the perturbation theory for $K(s|\,x,y)$ \cite{CPTII,nnea} --
there is no nonanalytic terms in $1/s$ like $\ln(1/s)$), one
easily obtains the series of recurrent equations for the
coefficients of this expansion. They start with
    \begin{eqnarray}
    &&F(\nabla)\,\Omega_0(x,y)=0,       \label{2.4}\\
    &&F(\nabla)\,\Omega_1(x,y)=
    \left(\sigma^\mu\nabla_\mu
    +\frac12\,\Box\,\sigma
    -\frac{d}2\,\right)\Omega_0(x,y).   \label{2.5}
    \end{eqnarray}

An obvious difficulty with the choice of their concrete solution
is that they do not form a well posed boundary value problem.
Indeed, natural zero boundary conditions at infinity for the
original kernel $K(s|\,x,y)$ do not impose any boundary conditions
on the function $\Omega(s|\,x,y)$ except the restriction on the
growth of $\Omega(s|\,x,y)$ to be slower than
$\exp\,[+\sigma(x,y)/2s]$ (in view of the exponential factor in
(\ref{2.1})). On the other hand, this freedom in choosing
non-decreasing at $|x|\to\infty$ solutions facilitates their
existence. In particular, the elliptic equation (\ref{2.4}) with
positive definite operator $F(\nabla)$ (which we assume) would not
have nontrivial solutions decaying at spacetime infinity. Thus,
the only remaining criterion for the selection of solutions in
(\ref{2.4})-(\ref{2.5}) is the requirement of their symmetry in
the arguments $x$ and $y$. As we will see now, this criterion
taken together with certain assumptions of {\em naturalness}
result in concrete solutions which will be further checked on
consistency by different methods including perturbation theory,
the variational equation for the heat kernel trace (\ref{1.2a})
and its metric analogue, etc.

The way this strategy works in the leading order of the $1/s$-expansion
was demonstrated in \cite{nnea} and is as follows. Make a natural
{\em assumption} that $\Omega_0(x,y)$ at $|x|\to\infty$ is not growing
and {\em independent} of the angular direction $n^\mu=x^\mu/|x|$
quantity $C(y)$ -- the function of only $y$. Then the solution of the
corresponding boundary value problem
    \begin{eqnarray}
    &&F(\nabla)\,\Omega_0(x,y)=0,      \nonumber\\
    &&\Omega_0(x,y)\,\Big|_{\,|x|\to\infty}=C(y),\nonumber
    \end{eqnarray}
is unique and reads $\Omega_0(x,y)=\Phi(x)\,C(y)$,
where $\Phi(x)$ is a special function (\ref{1.5}) solving
the homogeneous equation subject to unit
boundary conditions at infinity. Then, the requirement of symmetry in
$x$ and $y$ implies that $\Omega_0(x,y)=C\,\Phi(x)\,\Phi(y)$, where
the value of the numerical normalization coefficient $C=1$ follows
from the comparison with the exactly known heat kernel in flat spacetime
with vanishing potential $V(x)=0$. Thus
    \begin{eqnarray}
    \Omega_0(x,y)=\Phi(x)\,\Phi(y).      \label{2.6}
    \end{eqnarray}
This answer was checked in \cite{nnea} in few lowest orders of perturbation
theory in powers of the potential.

Substituting the expansion (\ref{1.4}) for the coincidence limit
$K(s\,|\,x,x)$ in the variational equation for ${\rm Tr}\,K(s)$
(\ref{1.2a}) one has the corresponding variational equation for
$W_0$ in Eq.(\ref{1.6}),
    \begin{eqnarray}
    \frac{\delta \,W_0}{\delta V(x)}=
    -g^{1/2}(x)\,\Omega_0(x,x)
    =-g^{1/2}(x)\Phi^2(x).    \label{2.13a}
    \end{eqnarray}
Its integrability --  the symmetry of the
variation of its right-hand side with respect to $V(y)$ in $x$ and
$y$ -- can be checked with the use of the following variational
derivative
    \begin{eqnarray}
    \frac{\delta\Phi(x)}{\delta V(y)}
    =G(x,y)\,\Phi(y)                  \label{2.14}
    \end{eqnarray}
which, in its turn, follows from the variation of the inverse
operator
    \begin{eqnarray}
    \delta\frac1{F(\nabla)}=
    -\frac1{F(\nabla)}\,\delta F(\nabla)\frac1{F(\nabla)}.
    \end{eqnarray}
Applying (\ref{2.14}) in the right-hand of (\ref{2.13a}) one finds
    \begin{eqnarray}
    \frac{\delta}{\delta
    V(y)}\,g^{1/2}(x)\,\Omega_0(x,x)
    =2g^{1/2}(x)\,\Phi(x)\,G(x,y)\,
    \Phi(y),                               \label{2.15}
    \end{eqnarray}
which is symmetric in $x$ and $y$ in view of the symmetry of
the Green's function\footnote{Which follows
from the hermiticity of the operator $F(\nabla)$ in the measure
$g^{1/2}(x)$ and the assumption that $G(x,y)$ is a density with
respect to $y$.},
    \begin{eqnarray}
    g^{1/2}(x)\,G(x,y)=g^{1/2}(y)\,G(y,x).  \label{sym}
    \end{eqnarray}

Thus, the equation is integrable and its explicit solution
(\ref{1.7}) can be checked by direct variation again with the use
of (\ref{2.14}),
    \begin{eqnarray}
    -\frac{\delta}{\delta
    V(y)}\int dx\,g^{1/2}(x)\,V\Phi(x)\,
    &=&\,-g^{1/2}(x)\,\Phi(y)\,\left(1+\int dx\,G(y,x)\,
    \Phi(x)\right)\nonumber                    \\
    &=&\,-g^{1/2}(y)\,\Omega_0(y,y).
    \end{eqnarray}

\section{Subleading order: particular case of flat spacetime}
\hspace{\parindent}
In the subleading order of $1/s$-expansion the situation is more
complicated. The next coefficient $\Omega_1(x,y)$ satisfies the
inhomogeneous equation (\ref{2.5}) the right hand side of which
can be rewritten in the form
    \begin{eqnarray}
    F(\nabla)\,\Omega_1(x,y)=
    \frac12\,[\,\stackrel{\rightarrow}{F}\!(\nabla_x)
    \Phi(x)\,\sigma(x,y)-d\,\Phi(x)\,]\,\Phi(y)  \label{2.7}
    \end{eqnarray}
in view of the equation for $\Phi$, $F(\nabla)\Phi=0$. A natural
solution $\Omega_1(x,y)=\psi(x,y)/2$ with
    \begin{eqnarray}
    \psi(x,y)=\frac1{F(\nabla_x)}
    \left[\,\stackrel{\rightarrow}{F}(\nabla_x)
    \Phi(x)\sigma(x,y)
    -d\Phi(x)\,\right]\Phi(y)                    \label{2.9}
    \end{eqnarray}
is not, however, correct because it violates the symmetry in $x$ and $y$.
Symmetric solution differs from this one by some solution of the
homogeneous equation. The latter can be obtained by projecting a rather
generic two-point function $v(x,y)$ onto the space of solutions by
the nonlocal projector $\Pi(\nabla_x)$
    \begin{eqnarray}
    \Pi(\nabla)=1-\frac1{F(\nabla)}\!
    \stackrel{\rightarrow}{F}\!(\nabla).   \label{projector}
    \end{eqnarray}
Here the arrow indicates the action of the differential
operator in the direction opposite to its Green's function,
$1/F(\nabla)$, written in the operator form. That is, the
action of this projector on $v(x,y)$ in
    \begin{eqnarray}
    \Omega_1(x,y)=\frac12\psi(x,y)
    +\Pi(\nabla_x)\,v(x,y)           \label{v}
    \end{eqnarray}
implies that
    \begin{eqnarray}
     \Pi(\nabla_x)\,v(x,y)=v(x,y)-
    \int dz\,G(x,z)\,
    \stackrel{\rightarrow}{F}\!(\nabla_z)\,v(z,y),
    \end{eqnarray}
and the integration by parts that would reverse the action
of $F(\nabla_z)$ on $G(x,z)$ (and, thus, would lead to a
complete cancellation of the first
term) is impossible without generating nontrivial surface terms.

The needed symmetry of $\Omega_1(x,y)$ can be attained by choosing
$v(x,y)=\psi(y,x)/2$ in (\ref{v}) such that the special solution
of the homogeneous equation takes the form
    \begin{eqnarray}
    \Pi(\nabla_x)\,v(x,y)=\frac12\,\psi(y,x)
    -\frac12\,\psi(y,x)
    \stackrel{\leftarrow}{F}\!(\nabla_x)
    \frac{\stackrel{\leftarrow}1}
    {F(\nabla_x)}                            \label{2.9a}
    \end{eqnarray}
(here we again use the operator notations for the Green's function
and the operator $\stackrel{\leftarrow}{F}\!(\nabla_x)$ acting,
this time, on $\psi(y,x)$ from the right).
Remarkably, in view of the structure of the function (\ref{2.9})
and the equation $F(\nabla)\Phi(x)=0$ the second term here turns
out to be symmetric in $x$ and $y$. Therefore by adding
(\ref{2.9a}) to the solution of the inhomogeneous equation
$\psi(x,y)$ we finally obtain the needed symmetry of
$\Omega_1(x,y)$
  \begin{eqnarray}
  &&\Omega_1(x,y)=\frac12\,\psi(x,y)+\frac12\psi(y,x)\nonumber\\
  &&\qquad\qquad\qquad\qquad
  -\frac12\,\frac1{F(\nabla_x)}
  \stackrel{\rightarrow}{F}\!(\nabla_x)\,
  [\,\Phi(x)\,\sigma(x,y)\,\Phi(y)\,]
  \stackrel{\leftarrow}{F}\!(\nabla_y)
  \frac{\stackrel{\leftarrow}{1}}{F(\nabla_y)}   \label{2.8}
  \end{eqnarray}

Interestingly, the analogue of the variational equation (\ref{2.13a})
for the subleading term of the $1/s$-expansion of ${\rm Tr}\,K(s)$
    \begin{eqnarray}
    \frac{\delta \,W_1}{\delta V(x)}=
    -g^{1/2}(x)\,\Omega_1(x,x)         \label{2.13}
    \end{eqnarray}
also satisfies the integrability condition and has a formal solution
in terms of the Green's function of $F(\nabla)$. As shown in Appendix A
it reads as
    \begin{eqnarray}
    W_1=\frac12\,\int dx\,dy\,g^{1/2}(y)\,
    \big[\stackrel{\rightarrow}{F}\!(\nabla_x)\,
    \Phi(x)\,\sigma(x,y)\,\Phi(y)
    \stackrel{\leftarrow}{F}\!(\nabla_y)\,
    \big]\,G(y,x),                           \label{1.8}
    \end{eqnarray}
where the operators in square brackets are acting in the directions
indicated by arrows on the arguments of
$\Phi(x)\,\sigma(x,y)\,\Phi(y)$.

Unfortunately, however, the validity of the algorithms (\ref{2.8})
and (\ref{1.8}) can at the moment be rigorously established only
in flat spacetime. Problem is that the nonlocal function
$\psi(x,y)$ is well (and uniquely) defined only when the
expression in square brackets of (\ref{2.9}) sufficiently rapidly
goes to zero at spacetime infinity. This expression has two terms
    \begin{eqnarray}
    &&\stackrel{\rightarrow}{F}(\nabla_x)\Phi(x)\,\sigma(x,y)
    -d\Phi(x)=2\sigma^\mu(x,y)\nabla_\mu\Phi(x)
    +\Phi(x)\,[\,\Box\,\sigma(x,y)-d\,].           \label{2.10}
    \end{eqnarray}
The first term has a power law falloff $1/|x|^{d-2}$ at
$|x|\to\infty$ in view of the behavior of $\sigma^\mu(x,y)\sim|x|$ and
$\nabla_\mu\Phi(x)\sim 1/|x|^{d-1}$. This
makes the contribution of this term (convolution with the kernel of
Green's function in (\ref{2.9})) well defined at least in dimensions
$d>4$. On the contrary, the second term is proportional to the
deviation of geodesics $\Box\,\sigma(x,y)-d$ which has
the following rather moderate falloff
    \begin{eqnarray}
      \Box\,\sigma(x,y)-d\sim\frac1{|x|},
      \,\,\,\,\, |x|\to\infty.               \label{2.10a}
    \end{eqnarray}
Therefore a purely metric contribution to (\ref{2.9}) turns out to be
quadratically divergent in the infrared. Tracing the origin of
this difficulty back
to the equation (\ref{2.5}) we see that the source term in its
right hand side is $O(1/|x|)$, so that the solution
$\Omega_1(x,y)\sim|x|$ is not vanishing at infinity and, therefore,
is not uniquely fixed by Dirichlet boundary conditions. Some
principles of fixing this ambiguity would
certainly regularize the integral in the definition of $\psi(x,y)$ and
uniquely specify all quantities in the subleading order. Unfortunately,
we do not have these principles at the moment. That is why in
what follows we will restrict the consideration of this
order to the flat-space case where this problem does not
arise at all.

In flat spacetime the geodesic deviation scalar (\ref{2.10a}) is
identically vanishing, because
    \begin{eqnarray}
    &&g_{\mu\nu}=\delta_{\mu\nu},\,\,
    \sigma(x,y)=\frac12\,|x-y|^2,\,\,\,
    \sigma^\mu(x,y)=(x-y)^\mu,\nonumber\\
    &&\Box\,\sigma(x,y)=d.                  \label{2.22}
    \end{eqnarray}
Therefore, the expression for $\psi(x,y)$ becomes well defined.
Correspondingly, in the square brackets of (\ref{1.8}) only one term
containing $\nabla^\mu_x\nabla^\nu_y\sigma(x,y)=-\delta^{\mu\nu}$
survives and yields
    $\stackrel{\rightarrow}{F}\!(\nabla_x)\,
    \Phi(x)\,\sigma(x,y)\,\Phi(y)
    \stackrel{\leftarrow}{F}\!(\nabla_y)=
    -4\,\nabla_\mu\Phi(x)\,\nabla^\mu\Phi(y)$,
so that $\Omega_1(x,y)$ and the subleading term of the functional
trace $W_1$ considerably simplify
    \begin{eqnarray}
    &&\Omega_1(x,y)=\frac1{\Box_x-V_x}\,(x-y)^\mu
    \nabla_\mu\Phi(x)\,\Phi(y)+(x\leftrightarrow y)\nonumber\\
    &&\qquad\qquad\qquad\qquad\qquad\qquad\qquad
    +\,2\,\frac1{\Box_x-V_x}\nabla_\mu\Phi(x)\,
    \frac1{\Box_y-V_y}
    \nabla^\mu\Phi(y),                       \label{2.24}
    \end{eqnarray}
    \begin{eqnarray}
    W_1=-2\,\int
    dx\,dy\,\nabla_\mu\Phi(x)\,\nabla^\mu\Phi(y)\,
    G(y,x)
    =-2\int dx\,\nabla_\mu\Phi
    \frac1{\Box-V}\nabla^\mu\Phi(x).         \label{2.25}
    \end{eqnarray}
The last expression coincides with the second term of (\ref{1.2})
obtained in \cite{nnea} by direct summation of perturbation series and,
thus, confirms the present nonperturbative (in the potential $V$)
method.

\section{Metric dependence}
\hspace{\parindent}
In this section we perform a major check on the validity of
the asymptotics (\ref{1.7}) in curved spacetime.
It is determined by its functional derivative
with respect to $V(x)$ only up to arbitrary metric functional.
This functional can be determined from the metric variational
derivative of ${\rm Tr}K(s)$ -- the analogue of Eq.(\ref{1.2a}).
So we derive the corresponding equation below and show that
the asymptotics (\ref{1.7}) indeed satisfies it, which confirms the
spacetime integral (bulk) part of (\ref{1.7}). Local variational
derivative $\delta/\delta g_{\mu\nu}(x)$ at finite $|x|$ cannot
probe possible surface integrals (of {\em local} combinations of
metric and its derivatives) at spacetime infinity, so that the
additional surface term (\ref{1.7a}) will be recovered in the
next section by another method.

From the operator definition of the heat kernel (\ref{1.1}) it follows
that its metric variation reads
    \begin{equation}
    \delta_g {\rm Tr}K(s)=-s{\rm Tr}\,
    \big(\delta_g F\,K(s)\big)
    =-s\int dx\,
    \delta_g F(\nabla_x)\,K(s|x,x')\,\Big|_{x'=x},  \label{3.1}
    \end{equation}
where the variation of the operator coincides with that of
the covariant d'Alem\-ber\-tian acting on scalars (\ref{1.3}) and
equals
  \begin{eqnarray}
  \delta_g F(\nabla)=\delta_g\Box=
  -\delta g_{\mu\nu}\nabla^\mu\nabla^\nu
  -\frac12(\nabla^\lambda\delta g_{\mu\nu})
  (\delta^\mu_\lambda\nabla^\nu
  +\delta^\nu_\lambda\nabla^\mu
  -g^{\mu\nu}\nabla_\lambda).                     \label{3.2}
  \end{eqnarray}
The corresponding variational derivative can be rewritten in
the form of the following integral bilinear in two test functions
$\varphi(x)$ and $\psi(x)$
  \begin{eqnarray}
  &&\int dx\,g^{1/2}\psi(x)\,
  \frac{\delta F(\nabla)}
  {\delta g_{\mu\nu}(y)}\,\varphi(x)
  =-g^{1/2}f^{\mu\nu}(\nabla_x,\nabla_y)\,
  \varphi(x)\,\psi(y)\,\Big|_{x=y}\nonumber\\
  &&\qquad\qquad\qquad\qquad\qquad\qquad=-g^{1/2}\psi(y)
  f^{\mu\nu}(\stackrel{\rightarrow}{\nabla}_y,
  \stackrel{\leftarrow}{\nabla}_y)\,
  \varphi(y).                         \label{3.3}
  \end{eqnarray}
The kernel of this local form is given by the differential operator
$f^{\mu\nu}(\nabla_x,\nabla_y)$ with covariant derivatives acting on two
different arguments $x$ and $y$ (or correspondingly to the right and to
the left as indicated above by arrows)
    \begin{eqnarray}
    f^{\mu\nu}(\nabla_x,\nabla_y)=
    -\nabla_x^{(\mu}\nabla_y^{\nu)}
    +\frac12\,g^{\mu\nu}\Box_x
    +\frac12\,g^{\mu\nu}
    \nabla^\lambda_x\nabla_\lambda^y.            \label{3.4}
    \end{eqnarray}
Using the expression (\ref{2.1}) for $K(s\,|\,x,y)$ and a simple
relation
$f^{\mu\nu}(\nabla_x,\nabla_y)\,\sigma(x,y)\,\big|_{y=x}=g^{\mu\nu}$
one has
    \begin{eqnarray}
    &&\frac{\delta{\rm Tr} K(s)}
    {\delta g_{\mu\nu}(x)}
    =-s\,g^{1/2}(x)\,f^{\mu\nu}(\nabla_x,\nabla_y)
    K(s|x,y)\,\Big|_{x=y}\nonumber\\
    &&\qquad\qquad=
    \frac{g^{1/2}(x)}{(4\pi s)^{d/2}}\,\left[
    \,\,\frac12 \,g^{\mu\nu}\,\Omega(s|\,x,x)
    -s\,f^{\mu\nu}(\nabla_x,\nabla_y)\,
    \Omega(s|\,x,y)\,\Big|_{\,y=x}\,\right],     \label{3.5}
    \end{eqnarray}
where the first term arises from the action of
$f^{\mu\nu}(\nabla_x,\nabla_y)$ on the exponential in
$K(s\,|\,x,y)$. Therefore the metric variational derivatives
of ${\rm Tr}\,K(s)$ in the first two orders
of the $1/s$-expansion become
    \begin{eqnarray}
    &&\frac{\delta W_0}{\delta g_{\mu\nu}}=
    -g^{1/2}\,f^{\mu\nu}(\nabla_x,\nabla_y)\,
    \Omega_0(x,y)\,\Big|_{\,y=x},             \label{3.6a}\\
    &&\frac{\delta W_1}{\delta g_{\mu\nu}}=
    \frac12\,g^{1/2}g^{\mu\nu}\Phi^2(x)-g^{1/2}
    f^{\mu\nu}(\nabla_x,\nabla_y)\,
    \Omega_1(x,y)\,\Big|_{\,y=x},              \label{3.6}
    \end{eqnarray}
where we took into account that $\Omega_0(x,x)=\Phi^2(x)$.

In the rest of this section we will focus at checking the
relation (\ref{3.6a}). For this purpose let us first calculate its
right hand side. After substituting the expression
$\Omega_0(x,y)=\Phi(x)\,\Phi(y)$ and taking into account the
relation $\Box\Phi=V\Phi$ one finds
  \begin{eqnarray}
  -f^{\mu\nu}(\nabla_x,\nabla_y)\,
  \Omega_0(x,y)\,\Big|_{\,y=x}
       =-\frac12\,\Big(g^{\mu\nu}V\Phi^2
  -2\nabla^\mu\Phi \nabla^\nu\Phi
  +g^{\mu\nu}\nabla^\lambda\Phi
  \nabla_\lambda\Phi\Big).                       \label{3.8}
  \end{eqnarray}

To find the metric variational derivative in the left hand side of
(\ref{3.6a}), we first note that the surface integral (\ref{1.7a})
does not contribute to it for any finite $|x|$. Then write down
the variational derivative of $\Phi(x)=\Phi(x)[\,g_{\mu\nu}\,]$.
The variation of the nonlocal Green's function in (\ref{1.5})
gives
  \begin{eqnarray}
  \frac{\delta\Phi(x)}{\delta g_{\mu\nu}(y)}=
  -\frac1{F(\nabla)}\,\frac{\delta F(\nabla)}
  {\delta g_{\mu\nu}(y)}\,\frac1{F(\nabla)}V(x)=
  -\frac1{F(\nabla)}\,\frac{\delta F(\nabla)}
  {\delta g_{\mu\nu}(y)}\left(\Phi(x)-1\right),   \label{3.9}
  \end{eqnarray}
where we used the relation $\Big(1/F(\nabla)\Big)V(x)=\Phi(x)-1$.
Then in view of the expression (\ref{3.3}) for
$\delta F(\nabla)/\delta g_{\mu\nu}$
  \begin{eqnarray}
  \frac{\delta\Phi(x)}{\delta g_{\mu\nu}(y)}=G(x,y)\,
  f^{\mu\nu}(\stackrel{\rightarrow}{\nabla}_y,
  \stackrel{\leftarrow}{\nabla}_y)
  \,\Phi(y).                                    \label{3.11}
  \end{eqnarray}

Let us integrate this equation over $x$ with the
(densitized) potential $g^{1/2}V(x)$. Then using the symmetry
of the Green's function of $F(\Box)$ one has
  \begin{eqnarray}
  \int dx\,g^{1/2}(x)\,V(x)\,\frac{\delta\Phi(x)}
       {\delta g_{\mu\nu}(y)}
       =g^{1/2}(y)\,\left(\Phi(y)-1\right)\,f^{\mu\nu}
  (\stackrel{\rightarrow}{\nabla}_y,
  \stackrel{\leftarrow}{\nabla}_y)\,\Phi(y),      \label{3.12}
  \end{eqnarray}
or in view of the expression for $f^{\mu\nu}
  (\stackrel{\rightarrow}{\nabla}_y,
  \stackrel{\leftarrow}{\nabla}_y)$
  \begin{eqnarray}
  &&\int dx\,g^{1/2}(x)\,V(x)\,\frac{\delta\Phi(x)}
       {\delta g_{\mu\nu}(y)}=
       \frac12\,g^{1/2}\Big(-g^{\mu\nu}V\Phi\nonumber\\
  &&\qquad\qquad\qquad\qquad\qquad\qquad
  +g^{\mu\nu}V\Phi^2
  -2\nabla^\mu\Phi \nabla^\nu\Phi
  +g^{\mu\nu}\nabla^\lambda\Phi
  \nabla_\lambda\Phi\Big)(y).     \label{3.13}
  \end{eqnarray}
Thus finally
    \begin{eqnarray}
    &&\frac\delta{\delta g_{\mu\nu}(y)}
    \int dx\,g^{1/2}\Big( -V\Phi\Big)\nonumber\\
    &&\qquad\qquad\qquad\qquad
    =-\frac12\,g^{1/2}\,
       \Big(g^{\mu\nu}V\Phi^2
  -2\nabla^\mu\Phi \nabla^\nu\Phi
  +g^{\mu\nu}\nabla^\lambda\Phi
  \nabla_\lambda\Phi\Big),                       \label{3.14}
    \end{eqnarray}
the first term in the right hand side of (\ref{3.13}) being
cancelled by the variation of $g^{1/2}$ in the integration
measure. Comparison with (\ref{3.8}) finally confirms the relation
(\ref{3.6a}).

\section{Comparison with perturbation theory}
\hspace{\parindent}
Comparison with perturbation theory in flat space has actually
been done in Sect. 4. There the leading and subleading orders of
${\rm Tr}\,K(s)$ were shown to coincide with those of (\ref{1.2}),
which in turn were obtained in \cite{nnea} by direct summation of
the covariant perturbation series in potential. Here we will make a
similar check for the metric part of $W_0$ and, in particular,
reveal the metric surface term (\ref{1.7a}).

The leading order of the $1/s$-expansion for ${\rm Tr}\,K(s)$ was
obtained up to cubic order in curvature and potential
$\Re=(V,R_{\mu\nu})$ in \cite{CPTIII,asymp}. For a scalar operator
(\ref{1.1}), (\ref{1.3}), it looks like
    \begin{eqnarray}
    {\rm Tr}\,K(s)&=& \frac{s}{(4\pi s)^{d/2}}\int dx\, g^{1/2}\,
    \left\{P-P\frac1{\Box}P
    +\frac13\,P\frac1{\Box} R
    -\frac16\,R_{\mu\nu}\frac1{\Box} R^{\mu\nu}
    +\frac1{18}\,R\frac1{\Box} R\right.
\nonumber\\&&\qquad\quad
       +\,P\left(\frac1{\Box}P\right)\frac1{\Box}P
       -\frac16\,R\left(\frac1{\Box}{P}\right)
       \frac1{\Box}P
       -\frac13\,P\left(\frac1{\Box}{P}\right)
       \frac1{\Box}R
\nonumber\\&&\qquad\quad
       +\,\frac1{36}\,P\left(\frac1{\Box} R\right)
       \frac1{\Box}R
       +\frac1{18}\,R\left(\frac1{\Box} R\right)
       \frac1{\Box}P
       -\frac1{216}\,R\left(\frac1{\Box} R\right)
       \frac1{\Box} R
\nonumber\\&&\qquad\quad
       +\,\frac1{12}\,R\left(\frac1{\Box}
R^{\mu\nu}\right)\frac1{\Box} R_{\mu\nu}
       -\frac16\,R^{\mu\nu}\left(\frac1{\Box}
R_{\mu\nu}\right)\frac1{\Box} R
       \nonumber\\&&\qquad\quad
       +\,\frac16\,\left(\frac1{\Box} R^{\alpha\beta}\right)
    \left(\nabla_\alpha\frac1{\Box} R\right)
    \nabla_\beta\frac1{\Box} R
    \nonumber\\
    &&\qquad\quad
    -\,\frac13\,\left(\nabla^\mu\frac1{\Box} R^{\nu\alpha}\right)
    \left(\nabla_\nu\frac1{\Box}
      R_{\mu\alpha}\right)\frac1{\Box} R
    \nonumber\\
    &&\qquad\quad
    -\left.\frac13\,\left(\frac1{\Box} R^{\mu\nu}\right)
    \left(\nabla_\mu\frac1{\Box}
      R^{\alpha\beta}\right)\nabla_\nu\frac1{\Box} R_{\alpha\beta}
 +{\rm O}[\,\Re^4\,]\right\}\nonumber\\
    &&\qquad\quad +\,{\rm O}\,\left(\frac1{s^{d/2}}\right),
    \,\,\,\, s\rightarrow\infty.                   \label{4.1}
\end{eqnarray}
Here $P(x)$ is the redefined potential term of the operator,
    \begin{eqnarray}
    P(x)=\frac16\,R(x)-V(x),          \label{4.2}
    \end{eqnarray}
and every Green's functions of the {\em covariant curved-space}
d'Alembertian $\Box=g^{\mu\nu}\nabla_\mu\nabla_\nu$, $1/\Box$, is
acting on the nearest curvature or a potential standing to the
right of it. The tensor nature of the Green's function is not
explicitly specified here by assuming that it is always determined
by the nature of the quantity acted upon by $1/\Box$. To clarify
how efficiently these condensed notations allow one to
simplify the presentation, we explicitly write as an example
one of the nonlocal factors above, $(1/\Box)R_{\mu\nu}(x)$.
Manifestly it reads as
    \begin{eqnarray}
    \frac1{\Box}R_{\mu\nu}(x)\equiv
    \int dy\,
    G_{\mu\nu}^{\;\;\;\;\alpha\beta}(x,y)\,R_{\alpha\beta}(y),
    \end{eqnarray}
where $G_{\mu\nu}^{\;\;\;\;\alpha\beta}(x,y)$,
$\Box_x G_{\mu\nu}^{\;\;\;\;\alpha\beta}(x,y)=
    \delta^{\alpha\beta}_{\mu\nu}\delta(x,y)$,
$G_{\mu\nu}^{\;\;\;\;\alpha\beta}(x,y)\,
\big|_{\,|x|\to\infty}=0$, is the Green's function of $\Box$
acting on a second-rank symmetric tensor field with zero boundary
conditions at infinity.

Using (\ref{4.2}) in (\ref{4.1}) one finds that the
leading order term of ${\rm Tr}\,K(s)$ consists of two parts --
one explicitly featuring only the original potential $V$ acted
upon by Green's functions of the {\em curved-space} $\Box$'s and
another purely metric one
    \begin{eqnarray}
    W_0\,&=&\,-\int
    dx\,g^{1/2}\,\left\{\,V+V\frac1\Box\,V+V\frac1\Box\,
    V\frac1\Box\,V+O\left(V^4\right)\right\}\nonumber\\
    &&+\;\frac16\int dx\,g^{1/2}\,\left\{\,R
    -\,R_{\mu\nu}\frac1{\Box} R^{\mu\nu}
    +\frac12\,R\frac1{\Box} R\right.\nonumber\\
    &&\qquad\qquad\qquad
       +\frac12\,R\left(\frac1{\Box}
R^{\mu\nu}\right)\frac1{\Box} R_{\mu\nu}
       -R^{\mu\nu}\left(\frac1{\Box}
R_{\mu\nu}\right)\frac1{\Box} R
       \nonumber\\
       &&\qquad\qquad\qquad
       +\left(\frac1{\Box} R^{\alpha\beta}\right)
    \left(\nabla_\alpha\frac1{\Box} R\right)
    \nabla_\beta\frac1{\Box} R\nonumber\\
    &&\qquad\qquad\qquad
       -2\,\left(\nabla^\mu\frac1{\Box} R^{\nu\alpha}\right)
\left(\nabla_\nu\frac1{\Box}
R_{\mu\alpha}\right)\frac1{\Box} R \nonumber\\
    &&\qquad\qquad\qquad
    -\left.2\,\left(\frac1{\Box} R^{\mu\nu}\right)
    \left(\nabla_\mu\frac1{\Box}
R^{\alpha\beta}\right)\nabla_\nu\frac1{\Box}
R_{\alpha\beta}
 +{\rm O}\,[\,R_{\mu\nu}^4\,]\,\right\}.            \label{4.3}
    \end{eqnarray}
When deriving this expression from (\ref{4.1}) we took into
account that
    \begin{eqnarray}
    \int dx\,g^{1/2}\,V(x)
    \left(\frac1\Box\,V(x)\right)^2
    =\int dx\,g^{1/2}\,V\frac1\Box\,
    V\frac1\Box\,V(x),                   \label{4.4}
    \end{eqnarray}
where now all Green's function $1/\Box$ are acting to the right.
The expression (\ref{4.3}) should now be compared with the
expansion of (\ref{1.7}) in powers of $\Re=(V,R_{\mu\nu})$.

The nonlocal expansion of $\Phi(x)$ in (\ref{1.7})
    \begin{eqnarray}
    \Phi(x)=1+\frac1\Box\,V(x)+\frac1\Box\,
    V\frac1\Box\,V(x)+O\left(V^3\right)        \label{4.5}
    \end{eqnarray}
obviously recovers the first integral of (\ref{4.3}) explicitly
containing only powers of potential with {\em metric-dependent}
nonlocalities. The second integral in (\ref{4.3}) of purely metric
nature seems to be completely missing in the expression
(\ref{1.7}) for $W_0$. We have to clarify why this term does not
violate the metric variational equation (\ref{3.6a}) that was
directly checked above.

A crucial observation is that this term is a topological invariant
independent of local metric variations in the interior of
spacetime -- exactly in this class of $\delta g_{\mu\nu}(x)$ the
functional derivative of (\ref{3.6a}) was calculated in Sect.5.
Direct expansion in powers of the metric perturbation
$h_{\mu\nu}$, $g_{\mu\nu}=\delta_{\mu\nu}+h_{\mu\nu}$, on
flat-space background in cartesian coordinates shows that this
term reduces to the surface integral at spacetime infinity. For
the class of asymptotically flat metrics with $h_{\mu\nu}(x)\sim
1/|x|^{d-2}$, $|x|\to\infty$, this surface integral is linear in
perturbations (contributions of higher powers of $h_{\mu\nu}$ to
this integral vanish) and involves only a {\em local} asymptotic
behavior of the metric $g^\infty_{\mu\nu}(x)=\delta_{\mu\nu}
+h_{\mu\nu}(x)\,\Big|_{\,|x|\to\infty}$,
    \begin{eqnarray}
    \int dx\,g^{1/2}\,&&\left\{\,R
    -\,R_{\mu\nu}\frac1{\Box} R^{\mu\nu}
    +\frac12\,R\frac1{\Box} R\right.\nonumber\\
       &&+\frac12\,R\left(\frac1{\Box}
    R^{\mu\nu}\right)\frac1{\Box} R_{\mu\nu}
       -R^{\mu\nu}\left(\frac1{\Box}
    R_{\mu\nu}\right)\frac1{\Box} R\nonumber\\
       &&
       +\left(\frac1{\Box} R^{\alpha\beta}\right)
    \left(\nabla_\alpha\frac1{\Box} R\right)
    \nabla_\beta\frac1{\Box} R\nonumber\\
    &&-2\,\left(\nabla^\mu\frac1{\Box} R^{\nu\alpha}\right)
    \left(\nabla_\nu\frac1{\Box}
    R_{\mu\alpha}\right)\frac1{\Box} R \nonumber\\
    &&
    -\left.2\,\left(\frac1{\Box} R^{\mu\nu}\right)
    \left(\nabla_\mu\frac1{\Box}
    R^{\alpha\beta}\right)\nabla_\nu\frac1{\Box}
    R_{\alpha\beta}
    +{\rm O}[\,R_{\mu\nu}^4\,]\,\right\}\nonumber\\
    &&\nonumber\\
    &&\qquad\qquad\qquad\qquad\qquad
    =\int\limits_{|x|\to\infty} d\sigma^\mu\,
    \big(\partial^\nu
    h_{\mu\nu}-\partial_\mu h\Big)
    \equiv\Sigma\,[\,g_\infty\,].        \label{4.6}
    \end{eqnarray}
Here $d\sigma^\mu$ is the surface element on the sphere of radius
$|x|\to\infty$, $\partial^\mu=\delta^{\mu\nu}\partial_\nu$ and
$h=\delta^{\mu\nu}h_{\mu\nu}$. Covariant way to check this
relation is to calculate the metric variation of this integral and
show that its integrand is the total divergence which yields the
surface term of the above type linear in $\delta
g_{\mu\nu}(x)=h_{\mu\nu}(x)$. Thus, the correct expression for
$W_0$ modified by the the metric functional integration
``constant'' $\Sigma\,[\,g_\infty\,]$ is indeed given by
Eqs.(\ref{1.7})-(\ref{1.7a}), and this constant does not
contribute to the metric variational derivative $\delta W_0/\delta
g_{\mu\nu}(x)$ at any finite $|x|$.

For asymptotically-flat metrics with a power-law falloff at
infinity $h_{\mu\nu}(x)\sim M/|x|^{d-2}$, $|x|\to\infty$, the
contribution of $\Sigma\,[\,g_\infty]$ is finite and nonvanishing.
For example, for $(d+1)$-dimensional Einstein action foliated by
asymptotically-flat $d$-dimensional spatial surfaces this surface
integral yields exactly the ADM energy $M$ of the gravitational
system. In a covariant form it can also be rewritten as a
Gibbons-Hawking term $S_{GH}\,[\,g\,]=\Sigma\,[\,g_\infty]$ -- the
double of the extrinsic curvature trace $K$ on the boundary (with
a properly subtracted infinite contribution of the flat-space
background) \cite{GH}
    \begin{eqnarray}
    \Sigma\,[\,g_\infty]=-2\int_\infty\!
    d^{d-1}\sigma\,\Big(g^{(d-1)}\Big)^{1/2}\,
    \Big(K-K_0\Big).                    \label{4.7}
    \end{eqnarray}
Thus, this is the surface integral of the {\em local} function of
the boundary metric and its normal derivative. The virtue of the
relation (\ref{4.6}) is that it expresses this surface integral in
the form of the spacetime (bulk) integral of the {\em nonlocal}
functional of the bulk metric. The latter does not explicitly
contain auxiliary structures like the vector field normal to the
boundary, though these structures are implicitly encoded in
boundary conditions for nonlocal operations in the bulk integrand
of (\ref{4.6}). It should be mentioned here that the nontrivial
equation (\ref{4.6}) enlarges the list of relations between
nonlocal invariants derived in \cite{basis}. The difference of
this relation from those of \cite{basis} is that it is an infinite
series in curvatures and forms a nonvanishing topological
invariant, while the relations of \cite{basis} are homogeneously
cubic in curvatures and hold only for low spacetime
dimensionalities $d<6$.

Note also, in passing, that the definition of the topological
invariant (\ref{4.6}) can be rewritten as the {\em nonlocal}
curvature expansion of the (Euclidean) Einstein-Hilbert action
\cite{nonlocal}. It is important that this expansion begins with
the {\em quadratic} order in the curvature
    \begin{eqnarray}
    &&-\int dx\,g^{1/2}\,R(g)-
    2\int_\infty\!
    d^{d-1}\sigma\,\Big(g^{(d-1)}\Big)^{1/2}\,
    \Big(K-K_0\Big)\nonumber\\
    &&\qquad\qquad\qquad\qquad\quad
    =\int dx\,g^{1/2}\,\left\{\,
    -\Big(R^{\mu\nu}
    -\frac12\,g^{\mu\nu}R\Big)\,\frac1{\Box}R_{\mu\nu}
    +{\rm O}\,[\,R_{\mu\nu}^3\,]\,\right\},     \label{4.8}
    \end{eqnarray}
and the corresponding quadratic form is linear in the Einstein
tensor -- the fact that was earlier observed, up to surface terms,
in \cite{nlbwa} (see Eq.(112) in this reference). This observation
can serve as a basis for covariantly consistent nonlocal
modifications of Einstein theory \cite{nonlocal} motivated by the
cosmological constant and cosmological acceleration problems
\cite{A-HDDG}.

To summarize this section, we conclude that perturbation theory
confirms, up to the local surface term, the nonperturbative
algorithm for the leading order of the $1/s$-expansion.
Apparently, this surface term can also be grasped by the
variational technique of Sect.5 which will be done
elsewhere\footnote{To attain the surface term in (\ref{1.7}) one
should remember that the variational equation (\ref{3.1}) is based
on the cyclic property of the operator product under the sign of
the functional trace. This, in turn, is equivalent to integration
by parts without extra surface terms. This property is violated in
the lowest (first) order in the curvature \cite{CPTII} which gives
rise to the surface term of (\ref{1.7}).}.

\section{Discussion: induced cosmological constant and nonlocal
effective action} \hspace{\parindent} Thus we have generalized the
heat-kernel asymptotics of  \cite{nnea} to curved
asymptotically-flat spacetimes. Together with the trivial
covariantization of the flat-space bulk integral (\ref{1.2}) this
generalization includes the Gibbons-Hawking surface integral
(\ref{1.7a}) of the extrinsic curvature of the boundary.

Apart from this integral the leading asymptotics vanishes in the
absence of the potential $V$ which encodes non-gravitational (or
matter) fields of the system. This has a simple qualitative
explanation. Pure gravity has two derivatives in the interaction
vertex, which improves its infrared behavior -- graviton
scattering amplitudes have no infrared divergences even despite
the massless nature of the field. This fully agrees with the
vanishing of the leading asymptotics of ${\rm Tr}\,K(s)$ for the
effects probing {\em local} geometry, providing better convergence
properties of the integral (\ref{1.000}) at $s\to\infty$.

The contribution of the Gibbons-Hawking term probes only global
quantities like the ADM energy defined by the integral over
infinitely remote boundary. Therefore, it seems to be robust
against ultraviolet structure of the theory and is likely to be
universal for a wide class of models independently of their
microscopic nature. Apparently, this serves as a justification for
the phenomenological long-distance modifications of gravity theory
motivated by the cosmological constant problem \cite{A-HDDG}. The
nonlocal representation of the Einstein-Hilbert action (\ref{4.8})
plays important role in such modifications because it underlies
the construction of their covariant actions \cite{nonlocal}.

These modifications might arise not only within braneworld
theories like GRS \cite{GRS} or DGP \cite{DGP} models. Rather,
they can be mediated by new nonperturbative nonlocal contributions
to the quantum effective action \cite{nnea}. In their turn, these
contributions originate from the infrared asymptotics of the above
type. As soon as the results of \cite{nnea} are generalized to
curved spacetime, these effects can be directly analyzed in
gravitational models of interest and are currently under study
\cite{Barwork}. In connection with this it is worth sketching
possible directions of the further research. Clearly, they
incorporate possible generalizations of the obtained results and
should provide closing the loopholes in our formalism above.

One important generalization consists in overstepping the limits
of the asymptotically-flat spacetime. The simplest thing to do is
to consider the asymptotically deSitter boundary conditions. On
the one hand, they are strongly motivated by the cosmological
acceleration phenomenon and, on the other hand, by the
dS/CFT-correspondence conjecture inspired from string theory
\cite{dS/CFT}. This generalization implies essential modification
of both perturbative and nonperturbative techniques for the heat
kernel, the generalization of the Gibbons-Hawking term to
asymptotically dS-spacetimes, etc. Another generalization concerns
the inclusion of fields of higher spins with the covariant
derivatives in the d'Alembertian involving not only the metric
connection but the gauge field connection as well.

Open issues include the modification due to possible violation
of geodesic convexity in curved spacetime and the extension to
higher orders of late time expansion. Interestingly, both
(seemingly different) issues might be related because they both
involve the geodesic deviation of Eq.(\ref{2.10a}). Possible
contribution of caustics, briefly discussed in Sect. 3 above,
might be important because it is likely to give qualitatively
new terms originating from summation over multiple geodesics
\cite{camporesi}.
These terms cannot be reached by perturbations in contrast
to the partial summation of perturbation series underlying our
present results.

On the other hand, higher orders of the $1/s$-expansion can be
important within the cosmological constant problem.
In particular, the subleading order $O(1/s^{d/2})$ incorporates
the cosmological term of the quantum effective action
(\ref{1.000}). Indeed, this term is expected to appear as a
covariantization of the third term ($\sim 1$) in
the flat-space asymptotics (\ref{1.2}) of ${\rm Tr}\,K(s)$,
    \begin{equation}
    \frac1{s^{d/2}}\int dx\,\times 1
    \to\frac1{s^{d/2}}\int dx\,g^{1/2}(x)\times 1.
    \end{equation}
Interestingly, it has the same form also in the limit of
$s\to 0$, determined by the first
coefficient $a_0(x,x)=1$ of the Schwinger-DeWitt expansion
\cite{DeWitt,nnea}. Via the integral (\ref{1.000}) it generates
the ultraviolet-divergent cosmological term
    \begin{eqnarray}
    \SGamma_\Lambda=\Lambda_\infty\int dx\,g^{1/2},\,\,\,\,
    \Lambda_\infty=
    -\frac1{2(4\pi)^{d/2}}\int_0^\infty
    \frac{ds}{s^{1+d/2}}.                      \label{7.1}
    \end{eqnarray}
In fact, this expression is also infrared divergent in the
coordinate sense -- the volume integral $\int dx\,g^{1/2}$ for
asymptotically-flat spacetime diverges at $|x|\to\infty$.

Of course, the abundance of divergences indicates that the
cosmological constant cannot consistently arise in
asymptotically-flat spacetime. The contribution (\ref{7.1}) in
{\em massless} theories does not carry any sensible physical
information and is cancelled due to a number of interrelated
mechanisms. First, its cancellation is guaranteed by the
contribution of the local path-integral measure to the effective
action, which annihilates strongest (volume) divergences under
appropriate regularization of the path integral \cite{Bern}.
Another mechanism is based on the use of the dimensional
regularization which puts to zero all power-like divergences.
Interestingly, in the latter case this happens due to exact
cancellation of the ultraviolet divergence of (\ref{7.1}) at $s=0$
against its infrared counterpart at $s\to\infty$. This may be
regarded as a well-known statement that the cosmological constant
problem is of both infrared and ultraviolet nature\footnote{In
dimensional regularization the integral (\ref{7.1}) is
analytically continued to the domain of $d$ where it is convergent
either at the lower (ultraviolet) or upper (infrared) limits. The
pole parts of these two complimentary divergences are opposite in
sign and cancel one another.}. All these mechanisms, however, stop
working for {\em massive} theories or for theories with
spontaneously broken symmetry, where the induced vacuum energy
presents a real hierarchy problem \cite{Weinberg}.

Preliminary results of Sect. 4 for $W_1$ allow one to look at the
above mechanisms from a somewhat different viewpoint. To begin
with, the cosmological term structure $\int dx\, g^{1/2}$ in ${\rm
Tr}\,K(s)$ behaves differently at late times and at $s\to 0$. In
contrast to the $s\to 0$ limit, this term is completely absent at
$s\to\infty$ -- the functional $W_1$ given by (\ref{1.8}) does not
contain the part of zeroth order in the curvature and potential
(indeed, in the absence of the potential the expression
(\ref{1.8}) is linear in $\Box_x\Box_y\sigma(x,y)$ and vanishes
for flat spacetime). The functional $W_1$ was recovered from the
variational derivative with respect to $V(x)$, Eq. (\ref{1.2a}),
so one could have expected that the cosmological term should have
been added to (\ref{1.8}) as a functional integration "constant".
But this is not the case, because $W_1$ exactly satisfies the
metric variational equation (\ref{3.6}) (this will be shown
elsewhere \cite{Barwork}).

On the other hand, it was mentioned that in view of the slow
falloff properties of the geodesic deviation (\ref{2.10a}), the
expression (\ref{1.8}) cannot be trusted beyond flat spacetime.
However, the fact that it formally passes a subtle check of
Eq.(\ref{3.6}) suggests that under certain regularization of the
divergent integrals the algorithm (\ref{1.8}) will survive the
transition to curved spacetime. In contrast to (\ref{1.2}), it
does not contain the cosmological term integral $\sim\int
dx\,g^{1/2}$. This does not, however, indicate major contradiction
with the covariant curvature expansion of
\cite{CPTII,CPTIII,asymp}, because this integral is formal and
infrared divergent, which reflects the continuity of the spectrum
of the operator $F(\nabla)$ in asymptotically-flat
spacetime\footnote{The lowest order of this expansion is
responsible for this infrared divergent integral, and the subtlety
in its treatment was clearly emphasized in \cite{CPTII}.
Reconsidering its contribution with a special emphasis on surface
integrals at spacetime infinity is currently under study both
within the perturbation theory and the nonperturbative technique
of the present paper \cite{Barwork}.}.

Altogether, this might qualitatively alter the mechanisms of
induced cosmological constant and, in particular, exclude exact
cancellation of its ultraviolet and infrared contributions
occurring in the dimensional regularization case above. This
alteration is likely to result in a nonlocal effective action of
the type suggested in \cite{nnea}. In fact, the origin of
nonlocality is similar to the nonlocal representation of the
Einstein action (\ref{4.8}) generated by the subtraction of the
linear in metric perturbation part of the bulk integral. Effective
subtraction due to $W_1$ is currently under study. We expect that
this might bring to light interesting interplay between the
cosmological constant problem and infrared asymptotics of the heat
kernel and nonlocal effective action.

\setcounter{section}{0}
\renewcommand{\theequation}{\Alph{section}.\arabic{equation}}
\renewcommand{\thesection}{Appendix \Alph{section}.}

\section{${\rm Tr}\, K(s)$ in subleading order}
\hspace{\parindent} The check of the integrability condition for
(\ref{2.13}) in the subleading order is based on (\ref{2.14}). It
gives
    \begin{eqnarray}
    &&\frac{\delta}{\delta
    V(y)}\,g^{1/2}(x)\,\Omega_1(x,x)=
    g^{1/2}(x)\,G(x,y)\,\psi(y,x)+
    (x\leftrightarrow y)\nonumber\\
    &&\qquad\qquad\qquad\qquad\quad
    -g^{1/2}(x)\Phi(x)\,
    \Big[\,G(x,y)\,\sigma(x,y)+
    d\,G^2(x,y)\,\Big]\,\Phi(y)\nonumber\\
    &&\qquad\qquad\qquad\qquad\quad
    -g^{1/2}(x)\,G(x,y)\,\frac1{F_x}
    \stackrel{\rightarrow}{F}_x
    [\,\Phi(x)\,\sigma(x,y)\,\Phi(y)\,]
    \stackrel{\leftarrow}{F}_y
    \frac{\stackrel{\leftarrow}{1}}{F_y},  \label{2.16}
    \end{eqnarray}
where for brevity we denoted by $F_x=F(\nabla_x)$ and
    \begin{eqnarray}
    G^2(x,y)\equiv
    \frac1{F(\nabla_x)}\,G(x,y)
    =\frac1{F^2(\nabla_x)}\,\delta(x,y).   \label{2.17}
    \end{eqnarray}
The symmetry of this expression in $x$ and $y$ guarantees the
existence of the solution $W_1$. Direct verification of this
solution given by Eq.(\ref{1.8}) looks as follows.

To begin with, the expression (\ref{1.8}) can be rewritten in the
form
    \begin{eqnarray}
    W_1=\frac12\,\int dx\,g^{1/2}(x)\,
    \frac1{F(\nabla_x)}
    \stackrel{\rightarrow}{F}\!(\nabla_x)\,
    \Phi(x)\,\sigma(x,y)\,\Phi(y)
    \stackrel{\leftarrow}{F}\!(\nabla_y)\,
    \Big|_{\,y=x},                             \label{2.18}
    \end{eqnarray}
where the Green's function is represented in the operator form as
$1/F(\nabla_x)$ acting on the $x$-argument of
$\stackrel{\rightarrow}{F}\!(\nabla_x)\,\Phi(x)\,\sigma(x,y)\,\Phi(y)
\stackrel{\leftarrow}{F}\!(\nabla_y)$ with a {\em subsequent}
identification of $y$ and $x$. Then its variational derivative
equals
    \begin{eqnarray}
    &&\frac{\delta \,W_1}{\delta V(x)}=
    -\Phi(x)\,\sigma(x,y)\,\Phi(y)
    \stackrel{\leftarrow}{F}_y\!
    \left.\frac{\stackrel{\leftarrow}{1}}{F_y}\,
    g^{1/2}\,\right|_{\,y=x}\nonumber\\
    &&\qquad\qquad
    +\int dy\, dz\,\stackrel{\rightarrow}{F}_z
    \big[\,G(z,x)\,\Phi(x)\,\sigma(z,y)\,
    \Phi(y)\,\big]
    \stackrel{\leftarrow}{F}_y\,g^{1/2}(y)\,
    G(y,z)\nonumber\\
    &&\qquad\qquad+\,\frac12\,\frac1{F_x}
    \stackrel{\rightarrow}{F}_x[\,\Phi(x)\,
    \sigma(x,y)\,\Phi(y)\,]
    \stackrel{\leftarrow}{F}_y\!
    \left.\frac{\stackrel{\leftarrow}{1}}{F_y}\,
    g^{1/2}\,\right|_{\,y=x},                  \label{2.19}
    \end{eqnarray}
where the first two terms arise from the variation of operators
$(F(\nabla_x),\,F(\nabla_y))$ and functions $(\Phi(x),\,\Phi(y))$
in Eq.(\ref{2.18}), while the third term corresponds to the
variation of the Green's function. In the second term one can
integrate by parts without extra surface terms so that
$F_z=F(\nabla_z)$ would act on $G(y,z)$, because
$\big[\,G(z,x)\,\Phi(x)\,\sigma(z,y)\,\Phi(y)\,\big]$ vanishes at
$|z|\to\infty$. This removes integration over $z$ and yields the
coincidence limit $F_y\sigma(y,z)|_{z=y}=d$,
    \begin{eqnarray}
    &&\int dy\, dz\,G(y,z)\stackrel{\rightarrow}{F}_z\!
    \big[\,G(z,x)\,\Phi(x)\,\sigma(z,y)\,\Phi(y)\,\big]\!
    \stackrel{\leftarrow}{F}_y\,g^{1/2}(y)\nonumber\\
    &&\qquad\qquad\qquad\quad=\int dy\,g^{1/2}(y)\,
    G(y,x)\,\Phi(x)\,\left[\,\sigma(z,y)\,\Phi(y)
    \stackrel{\leftarrow}{F}_y\right]_{\,z=y}\nonumber\\
    &&\qquad\qquad\qquad\quad
    =d\,g^{1/2}\Phi(x)\,
    \frac{\stackrel{\rightarrow}{1}}{F_x}\,\Phi(x).  \label{2.20}
    \end{eqnarray}
The action of the Green's function on $\Phi(x)$ here is not a well
defined operation because $\Phi(x)\to 1$
at infinity and the integral is infrared divergent. However, the
first term of (\ref{2.19}) is also divergent, and together with
(\ref{2.20}) it forms the expression $-g^{1/2}\psi(x,x)$ (which
is well-defined at least for a flat spacetime). Therefore,
   \begin{eqnarray}
    &&\frac{\delta \,W_1}{\delta V(x)}=-g^{1/2}\psi(x,x)
    +\frac12\,g^{1/2}\,\frac1{F_x}
    \stackrel{\rightarrow}{F}_x[\,\Phi(x)\,
    \sigma(x,y)\,\Phi(y)\,]
    \stackrel{\leftarrow}{F}_y\!
    \left.\frac{\stackrel{\leftarrow}{1}}{F_y}\,
    \right|_{\,y=x}\nonumber\\
    &&\qquad\qquad\qquad\qquad\qquad=
    -g^{1/2}\Omega_1(x,x),                        \label{2.21}
    \end{eqnarray}
which finally proves the needed equation (\ref{2.13}).

\section*{Acknowledgements}

A.O.B. is grateful for hospitality of the Physics Department of
LMU, Munich, where a major part of this work has been done under
the support of the grant SFB375. The work of A.O.B. was also
supported by the RFBR grant No 02-01-00930 and the LSS grant
No 1578.2003.2.  The work of D.V.N. was supported by the RFBR grant
No 02-02-17054, the LSS grant No 1578.2003.2 and by the Landau
Foundation. V.M. thanks SFB375 for support. The work of Yu.V.G.
was supported by NSERC of Canada.

\end{document}